# LEAST-SQUARES JOINT DIAGONALIZATION OF A MATRIX SET BY A CONGRUENCE TRANSFORMATION[*]

MARCO CONGEDO

*GIPSA-lab (Grenoble Images Parole Signal Automatique) : CNRS (Centre National de la Recherche Scientifique) - Université Joseph Fourier,  Université Pierre Mendès-France, Université Stendhal, Grenoble INP (Institut National Polytechnique).      961 rue de la Houille Blanche - 38402 Saint Martin d'Hères, FRANCE*

DINH-TUAN PHAM

*LJK (Laboratoire Jean Kuntzmann), – CNRS (Centre National de la Recherche Scientifique) - Grenoble INP (Institut National Polytechnique) - Université Joseph Fourier, 51 rue des Mathématiques - 38402 Saint Martin d'Hères, FRANCE*

The approximate joint diagonalization (AJD) is an important analytic tool at the base of numerous independent component analysis (ICA) and other blind source separation (BSS) methods, thus finding more and more applications in medical imaging analysis. In this work we present a new AJD algorithm named SDIAG (Spheric Diagonalization). It imposes no constraint either on the input matrices or on the joint diagonalizer to be estimated, thus it is very general. Whereas it is well grounded on the classical least-squares criterion, a new normalization reveals a very simple form of the solution matrix. Numerical simulations shown that the algorithm, named SDIAG (spheric diagonalization), behaves well as compared to state-of-the art AJD algorithms.

## 1. Introduction

In the following we will indicate matrices by upper case italic letters (*A*), matrix sets by bold upper case italic letters (***A***), vectors and indexes by lower-case italic letters (*a*), scalars by lower-case letters (a) and integers by upper-case letters (A). Notations $(\cdot)^T$, $(\cdot)^{-1}$ and $\|(\cdot)\|$ indicate transpose, inverse and Frobenius norm, respectively. *diag*($\cdot$) and *off*($\cdot$) returns a matrix comprising only the diagonal and off-diagonal elements of the argument, respectively. $\lambda(\cdot)$ indicates an eigenvalue of the argument. $e_j$ indicates a vector where the $j^{th}$ element equals one and the others equals zero and $E_{ji}$ the matrix where the *ji* element equals one and the others zero. Given a set of input matrices

[*] This Research has been partially funded by the French National Research Agency (ANR) within the National Network for Software Technologies (RNTL), project Open-ViBE (Open Platform for Virtual Brain Environments).

$$C = \{C_k : k=1...K, K>2\}, (1)$$

the *joint diagonalization* (JD) problem consists in finding a matrix *B* such that all K *congruence* transformations $B^T C_k B$ [11, p. 324] result in diagonal matrices [1-9]; we are given *C* and we seek a *B* such that

$$B^T C_k B = \Lambda_k, (2)$$

where all $\Lambda_k$ are diagonal. Note that if (2) is true for matrix *B*, then it is true also for any matrix of the group *BYP*, where *Y* is a diagonal matrix and *P* a permutation matrix. This simply shows that AJD is solved up to a sign, scaling and permutation indeterminacy, just like in ICA and BSS. In the *approximate joint diagonalization* (AJD) problem, which is of practical interest in engineering, the input matrices are statistical estimations, thus they are perturbed by finite sampling error plus noise. Then we may state the AJD problem as

$$B^T (C_k + N_k) B = \Lambda_k + B^T N_k B, (3)$$

where the $N_k$s are sampling error plus noise matrices. Hereafter we assume that *B*, $C_k$ and $\Lambda_k$ are all N-dimensional square matrices and that the $C_k$ matrices are symmetric (real case) or Hermitian (complex case). In the sequel we will treat the real case for simplicity, for the extension in the complex domain is straightforward. In this paper we present a new least-squares (LS) iterative algorithm for finding *B* in the exact case and an approximation of *B* in the approximate case. The basic idea involves the minimization of

$$\Im^{OFF}(B) = \sum_k \left\| Off(B^T C_k B) \right\|^2, (4)$$

while avoiding the trivial solution *B=0*. Several constraints may serve this purpose. For example:

$$B^T B = I \, (5) \text{ and } diag(B^T C_0 B) = I \, (6).$$

Constraint (5) assumes that *B* is orthogonal and has been considered in this fashion in [1]. The authors proposed a solution by Givens rotations. We are rather interested in the general case, i.e., finding non-orthogonal *B*s. For this purpose (6) has been proposed almost simultaneously in [6] and [8]. $C_0$ is any positive definite matrix, such as the data covariance matrix or *I*, boiling down to fixing the filter gain or the norm of the filter vectors to unity, respectively. The algorithm starts by finding a matrix $H$ such that $H^TC_0H=I$ and then iteratively applies transformations that minimize the criterion using successive eigenvector decompositions satisfying (6). In [7] a similar LS idea has been used to perform simultaneous joint diagonalization and zero-diagonalization on two matrix sets, an approach that suits time-frequency data expansions. Contrarily to (5), constrained (6) by itself cannot prevent degenerate solutions, thus a penalty term proportional to -log|det($B$)| has been added to (4) in [9]. However the resulting algorithm is slow, even more so than the (already slow) algorithms in [6-8].

## 2. Method

Denoting by $b_i^T$ the $i^{th}$ row vector of $B^T$ and by $b_i$ its transpose (the $i^{th}$ column vector of $B$) let us define

$$M_n = \sum_k \left( C_k b_n b_n^T C_k^T \right) \quad (7)$$

and

$$M = \sum_k \left( C_k BB^T C_k^T \right) = \sum_n M_n , \quad (8)$$

with $n=1\ldots N$ and $k=1\ldots K$. Note for future use that multiplying the left-hand side of (2) by its transpose and summing across the K matrices we obtain

$$\sum_k \left( B^T C_k BB^T C_k^T B \right) = B^T MB = \sum_k \Lambda_k^2 \quad (9)$$

From (9) it results that the diagonality of $M$ (8) is a necessary condition for (2) to be true. Notice that such condition is meaningful regardless the sign of the entries of the $\Lambda_k$s and the symmetry of the matrices in $C$. We have seen that the scaling of the columns of $B$ is arbitrary. Let us fix the scale such that

$$b_n^T M_n b_n = 1, \text{ for all } n=1...N \quad (10)$$

and consider the minimization of (4) under constraint (10). Since

$$\sum_k \left\| \text{Off}\left(B^T C_k B\right) \right\|^2 = \sum_k \left\| \left(B^T C_k B\right) \right\|^2 - \sum_k \left\| \text{Diag}\left(B^T C_k B\right) \right\|^2$$

we may write the off-criterion as

$$\Gamma^{\text{Off}}(B) = tr\left(B^T M B\right) - \sum_n \left(b_n^T M_n b_b\right). \quad (11)$$

The method of Lagrange multipliers leads us to minimize

$$tr\left(B^T M B\right) - \sum_n \delta_n \left(b_n^T M_n b_b\right),$$

where the $\delta_n$ values are adjusted in order to satisfy constraint (10). The first derivative of $\Gamma^{\text{Off}}(B)$ with respect to the element $ji$ of $B$ is

$$\begin{aligned}\frac{\partial \Gamma^{\text{off}}(B)}{\partial B_{ji}} &= \sum_k 2tr\left[B^T C_k B \left(E_{ji} C_k B + B^T C_k E_{ji}\right)\right] \\ &\quad - 2\delta_i \sum_k \left[\left(b_n^T C_k b_n\right)\left(b_n^T C_k e_j + e_j^T C_k b_n\right)\right] \\ &= \delta_i \sum_k \left[4tr\left(B^T C_k B B^T C_k E_{ji}\right) - 4\left(b_n^T C_k b_n b_n^T C_k e_j\right)\right]\end{aligned}.$$

This is the $ij$ element of matrix

$$4B^T M - 4\Delta \begin{bmatrix} b_1^T M_1 \\ \dots \\ b_N^T M_N \end{bmatrix},$$

where $\Delta$ is a diagonal matrix having diagonal elements $\delta_n$. The solution to our optimization problem (11) must therefore satisfy (equating the above to zero)

$$\begin{bmatrix} b_1^T M_1 \\ \dots \\ b_N^T M_N \end{bmatrix} = \Delta^{-1} B^T M.$$

But this is

$$M_n b_n = \delta_n^{-1} M b_n, \quad (12)$$

which is readily recognized as a *system of nested generalized eigenvalue-eigenvector decomposition for matrix pencils* ($M_n$, $M$). For each pencil $b_n$ is the eigenvector associated with the largest eigenvalue and it is normalized so to satisfy (10), which using (7) implies that each $b_n$ is normalized so that

$$\sum_k \left( b_n^T C_k b_n \right)^2 = 1.$$

Eq. (12) shows that the system of generalized eigenvalue-eigenvector decompositions is a stationary point for criterion (11), or equivalently of criterion (4). This result matches our intuition noting that

$$\sum_k \left\| Diag\left( B^T C_k B \right) \right\|^2 = \sum_{k=1}^{K} \sum_{n=1}^{N} \left( b_n^T C_k b_n \right)^2 = \sum_{n=1}^{N} b_n^T M_n b_n$$

and

$$\sum_k \left\| \left( B^T C_k B \right) \right\|^2 = \sum_{k=1}^{K} \sum_{i=1}^{N} \sum_{j=1}^{N} \left( b_i^T C_k b_j \right)^2 = \sum_{n=1}^{N} b_n^T M b_n,$$

thus the sought eigenvectors are directions maximizing the sum of the square of the diagonal elements of the input matrices with respect to the sum of the squares of all their elements.

The above maximization problem has not known closed-form solution, as we may expect for an AJD problem, however, following [7] we can proceed iteratively row-by-row with mutual restriction. The general scheme for the optimization of (11), which we name SDIAG (Spheric Diagonalization), is reported here below.

---

**SDIAG Iteration Scheme**

Initialize *B* by a non-singular clever guess or by *I* if no guess is available.

*While not* Convergence *do*

    Obtain $M_1...M_N$ by (7) and their sum $M$

    **(Sphering)**

    Find a matrix $H$ such that $H^T M H = I$

    **(Optimal Directions)**

    *For n*=1 to N do find the principal eigenvector $u_n$ of $H^T M_n H$

    Update *B* as $B \leftarrow HU$, where $U = [u_1...u_N]$

    Normalize *B* so that $\sum_k \left( b_n^T C_k b_n \right)^2 = 1$, for *n*=1 to N

*End While*

---

SDIAG has a number of properties, to which we now turn. The following theorem applies in general:

**Theorem 2.1**

If $H^T M H = I$ (after the sphering stage), *then* $\sum_{n=1}^{N} \sum_{i=1}^{N} \lambda_i \left( H^T M_n H \right) = N$, *that is, the sum of the* N *eigenvalues of the* N *matrices $H^T M_n H$ equals* N.

**Proof** : since the trace of a square matrix equal the sum of its eigenvalues, using (8) we have

$$\sum_{n=1}^{N} tr\left( H^T M_n H \right) = tr\left( H^T \sum_{n=1}^{N} M_n H \right) = tr\left( H^T M H \right) = tr(I) = N. \quad \square$$

The following two theorems apply only at the limit of *B* in the exact JD case, that is, when the stationary point has been reached and (2) is true:

**Theorem 2.2**

*If $H^TMH = I$, let $u_n$ be the unit L2 norm eigenvector associated with the largest eigenvalue of $H^T M_n H$, i.e., $u_n^T H^T M_n H u_n = \lambda_1 \left( H^T M_n H \right)$, then square $U = [u_1 ... u_N]$ is orthogonal.*

**Proof**: Let us write $b_n = H u_n$. Eq. (9) states that $B^T M B = U^T H^T M H U$ is diagonal, thus for all $i, j=1...N$, $i \neq j$, we have $b_i^T M b_j = 0$, but since $H^TMH = I$ we have $u_i^T u_j = 0$, implying orthogonality of $U$ whether its vectors have unit L2 norm. □

**Theorem 2.3**

*If $H^TMH = I$, then for each $n=1...N$*
$$\begin{cases} \lambda_1 \left( H^T M_n H \right) = 1.0 \\ \lambda_2 \left( H^T M_n H \right), ..., \lambda_N \left( H^T M_n H \right) = 0.0 \end{cases},$$

*that is, the largest eigenvalue of all the N matrices $H^T M_n H$ equals 1.0 and all the others N-1 eigenvalues are null.*

**Proof**: Let $q_{kn} = H^T C_k H u_n$. For the $n^{th}$ matrix $M_n$ use (7) and $b_n = H u_n$ to write

$$H^T M_n H = \sum_{k=1}^{K} \left( H^T C_k H u_n u_n^T H^T C_k^T H \right) = \sum_{k=1}^{K} \left( q_{kn} q_{kn}^T \right).$$

Such matrix is a rank-1 projection matrix [11, p. 156], thus it has one non-null eigenvalue. Furthermore, the eigenvalues of a projection matrix are either 1.0 or 0 [11, p. 238], which proves the theorem for any given $n$. □

We end up this section with three remarks:

1. Theorems 2.1 to 2.3 above assume the existence of a matrix $H$ such that $H^TMH=I$. Being $M$ symmetric regardless the symmetry of $C$, such a matrix always exists if $M$ is positive definite. If this is not the case it suffices to reduce appropriately the number of columns of $H$, say, keeping only R<N of them; $H$ will be now of dimension NxR while R-dimensional $H^TMH=I$ will be satisfied. Finally, $U$ will still be square, but R-dimensional as well. Ideally, R should match the rank of $M$. In practice, we shall define R as the number of eigenvalues of $M$ larger then $\lambda_{max}(M)/f$, where f is an estimation of the ratio between the variance of the signal and the variance of the noise (by default we may set f=100).

2. The SDIAG algorithm naturally avoids degenerate solutions (see [9]). Referring to the previous point, $H$ has row-rank R≤N. Since at the stationary point $U$ is orthogonal (theorem 2.2) and multiplication by a nonsingular matrix does not alter rank [12, p. 197], $B^T$ will have raw-rank R as well. Its pseudo-inverse, the estimation of the "mixing" matrix in ICA/BSS applications, will have column-rank R and it will be a right inverse of $B^T$ [12, p. 199].

3. The N eigenvectors $u_n$ (optimal directions) sought at each iteration of SDIAG can be found in parallel running N threads. Furthermore, they can be found by power iterations (difference equations) of the form $u_n(p+1)=(M^{-1}M_n)u_n(p)$, where $p$ is the power iteration index [11, p. 359]. To avoid confusion with SDIAG iterations hereafter we name a power iteration a "pass". Note that theorem 2.3 ensures the stability of SDIAG, for the eigenvectors sought in the optimal directions step have eigenvalues bounded superiorly by 1.0 (neutrally stable as per [11, p. 259]). Now, If $\lambda_{n1}$ is the largest eigenvalue of $H^T M_n H$ associated to $u_n$ and $\lambda_{n2}$ is the next largest eigenvalue, power passes have convergence factor $\lambda_{n2}/\lambda_{n1}$ [11, p. 360]. From theorem 2.3 we see that the convergence factor approaches zero as the SDIAG algorithm converges, thus in the proximity of the solution the difference equations will convergence with only one pass.

## 3. Results

We compare our SDIAG algorithm to the well-established FFDIAG algorithm of [5] and QDIAG of [6]. Referring to (6), we use the sum of the input matrices as $C_0$ for QDIAG. We generate square diagonal matrices with each diagonal entry distributed as a chi-squares random variable with one degree of freedom. Each of these matrices, named $D_k$, may represent the error-free covariance matrix of independent standard Gaussian processes (zero mean and unit variance). The noisy input matrices are obtained following the noisy model in (3) such as

$$C_k + N_k, \ C_k = \left(AD_k A^T\right). \quad (13)$$

In (13), symmetric noise matrix $N$ has entries randomly distributed as a Gaussian with zero mean and σ standard deviation. The parameter σ controls the signal to noise ratio of the input matrices. Two different values will be considered in the simulations, of which one (σ=0.01) represents a small amount of noise closely simulating the exact JD case and the other (σ=0.03) simulating the approximate (realistic) JD case. Two kinds of mixing matrix *A* are considered in (13). In the general case mixing matrix *A* is obtained as the pseudo-inverse of a matrix with unit norm row vectors which entries are randomly distributed as a standard Gaussian; in this case (*A* non-orthogonal) the mixing matrix may be badly conditioned and we can evaluate the robustness of the AJD algorithms with respect to the conditioning of the mixing matrix. We also consider the case in which *A* is a random orthogonal matrix; in this case the conditioning does not jeopardize the performance of the algorithms and we can evaluate their robustness with respect to noise. As it is well known, given true mixing *A*, each AJD algorithm estimates demixing matrix $B^T$, which should approximate the inverse of actual *A* out of row scaling (including sign) and permutation. Then, matrix $G = B^T A$ should equal a scaled permutation matrix. At each repetition we compute the performance index as

$$\text{Perfirmance Index} = \frac{2(N\text{-}1)\sum_i \sum_j G_{ij}^2}{\sum_i \max_j \left(G_{ij}^2\right) + \sum_j \max_i \left(G_{ij}^2\right)}, \quad (14)$$

where *i* and *j* are the row and column index, respectively. Performance index (14) is positive and reaches its maximum 1.0 iff *G* has only one non-null elements in each row and column, i.e., if $B^T$ has been estimated exactly out of usual row scaling and permutation arbitrariness. The means and standard deviations obtained across 250 repetitions for 30 input matrices of dimension 10x10 are reported in table 1.

All pair-wise statistical tests between the mean performance of the three methods (bi-directional unpaired student-t with 248 degrees of freedom) reveal that the performance of the three algorithms is statistically equivalent in all conditions but in the case of non-orthogonal mixing and high noise ($\sigma=0.03$). In this condition the performance of FFDIAG is statistically lower then both QDIAG and SDIAG. This indicates that in noisy conditions FFDIAG occasionally fails in estimating correctly the demixing matrix *B* due to the ill-conditioning of the mixing matrix. In [10] we have performed simulations accounting for other kinds of perturbations of the exact JD model in (2). Those simulations showed that SDIAG is more robust than QDIAG to such violations. We note by the way that in [10] the correct criterion and consequent correct normalization for vectors $b_n$ (10) were not identified. Further work is currently in progress to study the convergence properties of SDIAG and an efficient implementation.

Table 1. Mean and standard deviation (in parentheses) of the performance index (14) attained by QDIAG [6], FFDIAG [5] and our SDIAG algorithm across 250 repetitions of the simulation with N=10 and K=30. The higher the mean the better the performance. See text for details.

|  | Orthogonal Mixing (good conditioning) | | Non-Orthogonal Mixing (variable conditioning) | |
| --- | --- | --- | --- | --- |
|  | $\sigma=0.01$ | $\sigma=0.03$ | $\sigma=0.01$ | $\sigma=0.03$ |
| QDIAG | 0.99978693 (0.00014084) | 0.99459649 (0.00804553) | 0.99978692 (0.00014071) | 0.99429150 (0.00885019) |
| FFDIAG | 0.99977825 (0.00015045) | 0.99555599 (0.00628480) | 0.99568611 (0.02840985) | 0.98832193 (0.03840436) |
| SDIAG | 0.99978186 (0.00014960) | 0.99539675 (0.00656474) | 0.99978183 (0.00014947) | 0.99521559 (0.00704235) |

## Acknowledgments

The author wish to express its gratitude to Prof. Christian Jutten, to Dr. Reza Sameni, Dr. Fabian Theis and Cédric Gouy-Pailler for the discussions about the ideas behind SDIAG.